\documentclass{iau}
\usepackage{graphicx}

\title[Conway's Game of Life: Livingness Beyond the Biological] 
{Conway’s Game of Life as an Analogue to a Habitable World: Livingness beyond the Biological}

\author[James McCrum \& Terence P. Kee ]   
{James McCrum $^1$
 \and Terence P. Kee$^2$}

\affiliation{$^1$School of Physics and Astronomy, University of St Andrews, St Andrews, UK \\ email: {\tt jm441@st-andrews.ac.uk} \\[\affilskip]
$^2$ School of Chemistry, University of Leeds, Leeds, UK  \\email: {\tt t.p.kee@leeds.ac.uk}}

\pubyear{2024}
\volume{387}  
\pagerange{119--126}
\jname{(Toward) Discovery of Life Beyond Earth and its Impact}
\editors{Hermine Landt, Martin Dominik \& Carol Oliver, eds.}
\begin{document}

\maketitle

\begin{abstract}
Conway’s Game of Life is a cellular automaton noted for its rich, complex, and emergent behaviour, which seems qualitatively 'lifelike'. It exists within a wider space of different rulesets of cellular automata, none of which have been found to display behaviours that seem as rich as Conway’s selected example. We present here a set of three quantitative tests for 'lifelike' behaviour, based on the critical brain theory, Shannon's theory of information entropy and integrated information theory, all of which are successfully able to select Conway's Game of Life as an outlier within this set, which is a non-biological analogue to the selection of a habitable planet or universe amongst a wider space of similar settings that cannot support the same kinds of living systems.

\end{abstract}

\firstsection 
\section{Introduction}

Life is an exceptionally difficult concept to pin down with specificity and validity. The exact parameters of what constitutes life and its distinctive behaviours and attributes are subject to immense debate and controversy, and represent one of the most important  unsolved problems in the interdisciplinary sciences. (\cite[Cleland \& Cheba. 2010]{ref1};\cite[Cornish-Bowden \& Cárdenas. 2019]{ref2}; \cite[McKay et al. 2004]{ref-c})

In light of this, those trying to solve this problem may wish to start with a simple representation of a lifelike system in order to try and build up from first principles what we mean when we say that a behaviour is lifelike, and therefore gain some insight into what we mean when we talk, in general, about life. 

John Von Neumann and others created the most basic and influential way to represent a simplistic ‘lifeform’;  cellular automata, a broad set of variable and versatile programs. (\cite[von Neumann \& Burks. 1966]{ref3})

'Automata' here refers to the idea that the program acts as a simulacrum of a biological organism, without itself being biological. Cellular refers to the fact that the particular feature of organic systems that the structure of the program emulates is modularity.

One may think of how a neural network creates a simplified automaton of a brain by selecting out of its many possible emulatable characteristics the specific characteristic of network structure and emulating this; a cellular automaton is similar in that it essentially reduces life to inter-cellular interactions, and emulates only this structure. 

Each cell has some finite number of discrete states as well as some finite set of neighbours to which it connects, and evolves in discrete generations according to its own state and the states of its neighbours. 

This is an absolutely bare-bones representation of how a living system functions, and yet they can be incredibly complex and display emergent behaviour that appears subjectively lifelike.

Even more striking than the initial cellular automata of the 1940s – which are relatively complicated in structure – is Conway’s Game of Life (\cite[Gardner. 1970]{ref4}), developed in the 1970s by Cambridge student John Conway (\cite[Izhikevich et al. 2015]{ref5}), who would go on to become one of the most influential mathematicians of the late 20th century, with many further accomplishments beyond this.

Conway’s Game of Life (CGOL) is striking because it is simultaneously one of the simplest cellular automata in structure and yet, one of the most complex in behaviour.  (\cite[Berkelamp et al. 2004]{ref6}) (\cite[Schiff. 2011]{ref7}). As the name suggests, CGOL  is exceptionally subjectively lifelike, and yet the basic essence of the program can be boiled down to two short binary strings. 

Each cell has two states, ‘dead’ or ‘alive’, so functions as a bit. Each cell carries out a simple logical operation on its 8 neighbours on the simple 2D grid it occupies – referred to as its Moore Neighbourhood - and then reacts based on its own state. All it does is count how many of its neighbours are alive. If there are two or three live neighbours and it is alive, it remains alive. If there are three neighbours and it is dead, it becomes alive. In any other case it will be dead in the next generation. 

This is relatively wordy spelled out in English, but if we write out two binary strings, one for a live cell and one for a dead cell, where the nth bit corresponds to n live neighbours and the state of that bit is 1 if this produces a live cell and 0 if it produces a dead cell, it reduces to: $$(0,0,1,1,0,0,0,0,0)$$ $$(0,0,0,1,0,0,0,0,0)$$ 
In essence you have an incredibly concisely defined artificial ‘genotype’ that nonetheless produces a shockingly elaborate phenotype. 

We can certainly say that in CGOL we have located a very simplistic representation of a life-like system. We can now ask what exactly it has in common with biological life - what quantitative traits produce its qualitative 'livingness'?

One core aspect of CGOL is that it is highly unpredictable. It is in general difficult to predict how patterns will evolve, which gives the system some sense of agency and independence. 
The first way to parametrise this unpredictability would be to say that the system is displaying \textbf{criticality}.

Criticality refers to the behaviour of a system very near a phase transition, such that small perturbations in the parameter(s) controlling the phase will shift the system to and from different phases, resulting in unpredictable behaviour. Familiar examples can be found in thermodynamics and in nuclear physics, but also in biology – the human brain is a critical system, and it has in fact been hypothesised that such criticality is essential to its proper functioning. (\cite[Tian et al. 2022]{ref8})  Furthermore, a key marker of criticality is fractality and self-similarity, which is also a key feature of many biological systems. (;\cite[Azua-Bustos \& Martínez. 2013 ]{ref-a};\cite[Grosu et al. 2023]{ref-b})

Like living things, CGOL is critical, in that very small changes to an initial condition on the board can lead to wildly divergent results; it also displays the corresponding ‘organic’ fractal patterns. (\cite[Alstrøm \& Leão. 1994]{ref10}) (\cite[Bak et al. 1989]{ref11})

Strikingly, CGOL’s criticality directly parallels the brain. The brain alternates between phases where neural firing is amplified, leading to propagation and an explosion in the number of firing cells, and where it is ignored, leading to decay and a population collapse after a timelapse. (\cite[Tian et al. 2022]{ref8}) CGOL alternates between phases where an initial small pattern of living cells will grow to cover much of the board, or will disintegrate into nothingness, or will stabilise in size. (\cite[Wolfram. 1984]{ref12})

The second way to discuss CGOL’s wild unpredictability is that it is \textbf{computationally irreducible}.

Computational irreducibility is a concept developed by Wolfram which, unlike criticality, was first derived from cellular automata like CGOL and then applied to biological systems like the brain. (\cite[Wolfram. 2002]{ref13}) A computationally irreducible system is one that is unable to be approximated by a simpler system, so cannot be ‘shortcut’, but can nonetheless be deterministic in nature. 

CGOL's computational irreducibility is a byproduct of its \textbf{Turing Completeness}.

The Turing Completeness property of CGOL is that one can build any computable program as a configuration in CGOL and then run said program by allowing the game to evolve from that configuration. If one could, given only the state of generation 1 of a game, predict the state of generation 10,000  one would be able in general to determine the outcome of such a program running. Turing showed that this – the Halting Problem – is undecidable. (\cite[Turing. 1937]{ref14}) Therefore one cannot 'shortcut' the system - the only way to find the state of generation 10,000 from generation 1 is to evolve all of the intervening states in CGOL through.

Similarly, Wolfram argued that the brain is computationally irreducible, in that one could know the exact state of a brain at a given time, but be unable to predict its state some time later using any simple algorithm or rule. Even if the brain, he argues, evolves deterministically according to some exact computation like CGOL, the only way to find the state is to let it evolve through all the in-between states, much as one cannot ‘skip’ states in CGOL. He presents this as an argument for free will without non-determinism.

What’s doubly striking here is that Conway located his specific eponymous game within a larger ‘multiverse’ of different games, not all of which are necessarily lifelike in this way, and that the reasoning he used to do was deeply tied to both of these models of its behaviour. Conway, as a Cambridge student, was interested in games and cellular automata, and wished to create a game that would be ‘universal’ – that is, Turing complete, able to simulate any program within its ruleset – and also unpredictable, in that it would not be trivial to determine how any given starting configuration will generally play out. He noted that he devised many prototype games over the course of 18 months, and discarded them all. He found them uninteresting, because they were highly predictable – initial patterns tended to either very predictably grow, or very predictably disintegrate. (\cite[Izhikevich et al. 2015]{ref5})

Although he did not phrase it in these terms, what Conway was searching for here was a critical system – one that is not in the phase of ‘population decay’ or ‘population growth’, but at an unpredictable liminal point between these. He eventually found this, and also a Turing complete system - another related desire of his for his game - in ‘Life’. (See fig 1a).

Although we cannot know what all of the rulesets Conway tested exactly were, we can easily define a space of rulesets that are similarly simple and adjacent to CGOL by simply modifying the population values in the Moore Neighbourhood that produce life or death. In practice that means that any two pair of nine-length binary strings can define a ruleset, which gives $2^{18}$ possible configurations. These are referred to as Lifelike Cellular Automata (LCA). 

Now that we have defined a very simple representation of a living system, and a number of directly comparable representations that are not as strongly associated with life, we can ask how, exactly, CGOL objectively differs, in order to gain insight into how biological life differs from abiotic systems.

\begin{figure}[h!]
\begin{center}
 \includegraphics[width=5                                                                                                                   in]{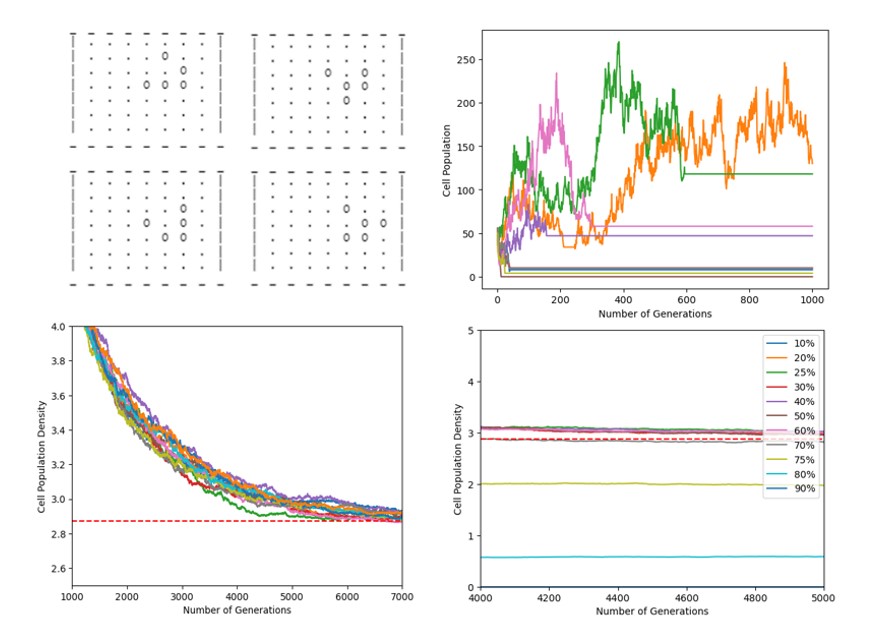} 
\caption{a) the 'Glider', a simple pattern in CGOL that carries information across the board, allowing for Turing Completeness. b) A set of random starting configurations in CGOL allowed to freely evolve; the difficulty in predicting when a given configuration will die out is a collorary of the Halting Problem.
 c) A set of  maximally entropic 'soups' of 50\% life probability in CGOL converging to an equilibrium; Flammenkamp's value of this is shown in red.
 d) The convergence of all but the most heavily populated soups to the Flammenkamp value.
} 
\label{fig1}
\end{center}
\end{figure}

\section{Methodology}
We here define three objective tests to which we submit any LCA in order to compare it to CGOL.

\vspace{3mm} Test 1) Given some set of initial starting conditions, what is the standard deviation of final population size compared to that of CGOL?

\vspace{3mm} Test 2) Given a maximally disordered starting system, what is the disorder of the final system?

\vspace{3mm} Test 3)	What is the integration parameter, $\Phi$, of this system?

\vspace{3mm} We already know from Conway’s own anecdotal account (\cite[Izhikevich et al. 2015]{ref5}) that many cellular automata – although not necessarily LCAs – have highly similar population values from a wide variety of different starting conditions, unlike CGOL, which is highly divergent from even minor variations in starting condition. So the first test serves as an easy way to compare the criticality/computational irreducibility of a system relative to CGOL. Starting from some set of different starting configurations, how distinct are the final population sizes after some number of generations? (See fig 1b).

The second test is based on the basic observation that biological systems are highly self-organising and tend to exist far from thermodynamic equilibrium; the idea that a defining separation between living and non-living things is ‘anti-entropy’ was popularised by Schrodinger in the 1940s, and so we can use it as another comparison of CGOL to its close counterparts in livingness. (\cite[Schrodinger. 1944]{ref15})

For this test, we consider that a discrete grid system like CGOL has the definable macrostate parameter of overall population of living cells, with the microstates being the specific cells that are alive or dead. The maximally disordered state with respect to this parameter would be a system where each cell has a 50\% chance of being alive or dead. (\cite[Macii \& Poncino. 1996]{ref16}) (\cite[Landauer. 1961]{ref17}) (\cite[Shannon. 1948]{ref18}) 

In Life-enthusiast parlance, a setup where each cell shares the same probability of being a 'live' cell is called a ‘soup’. It is already known that CGOL has a self-organising effect on soups of a wide range of probabilities, such that the grid drifts towards a population value with a much lower entropy over time. (\cite[Eppstein. 2010]{ref19}) (\cite[Flammenkamp. 2004]{ref20}) (See appendix 1, and fig 1c and 1d). This is, like unpredictability, a strikingly lifelike behaviour. So we can ask if other LCAs also do this.

The third test is based on a new application of the neuroscience idea of Integrated Information Theory (IIT). (\cite[Tononi et al. 2016]{ref21})

We have already seen that CGOL displays brain-like characteristics, such as criticality, fractality, and emergent complexity. We therefore turn to IIT, which provides a means of parametrising the resemblance to the brain of some non-organic system. 

As initially conceived, IIT was intended to measure consciousness, which it equates to the qualia of self-awareness. This is a highly controversial and convincingly disputed theory. (\cite[Doerig et al. 2019]{ref22}) (\cite[Merker et al. 2022]{ref23}) 

It may, however,  be possible to reinterpret $\Phi$, the ‘consciousness parameter’, as a ‘livingness parameter’, if we reason that systems that are highly self-connected will tend to more closely resemble living systems, as these systems are able to self-regulate and to recover from damage more effectively in a manner selected for by evolution.  This is by far the most speculative of the three tests. $\Phi$ can be seen as a measurement of the extent to which a system is indivisible, such that it cannot be defined merely as the sum of its parts. We test it here as a measurement of lifelike behaviour, using the PyPhi implementation. (\cite[Mayner et al. 2018]{ref24}) 

Given that we already know that CGOL was selected by a human observer – Conway – as especially lifelike and interesting out of the set of different cellular automata, if these tests are able to correctly isolate CGOL as an outlier from the broader LCA set, then they can be said to be able to successfully parametrise the human intuition of ‘livingness’.

\section{Sampling}
100 of the $2^{18}$  different LCAs were randomly selected by independently randomly generating binary strings in the range from 0 to $2^9$ for the living and dead cell rulesets. 

There are a number of different conventions in the literature used to refer to cellular automata of the lifelike class. Here, we use a novel one in which each cellular automaton is specified uniquely by a pair of decimal numbers between 0 and $2^9$, where the two numbers are the decimal representations of their binary rules trings. For example, CGOL is written 96/32, which is just the decimal representation of 001100000 / 000100000.

The full list of 100 LCAs used in this study for the first two tests, named in this notation, is given in Appendix 2. 

Each LCA was uniquely defined in two independent ways; on a box grid,  where cells on, for example, the top left edge of the grid, have no neighbours above or to the left; and on a toroidal grid, where cells on the top left edge of the grid neighbours cells on the bottom right edge of the grid. A toroidal grid wraps back on itself and maps to a torus – a box grid maps to a chess or Go board (the latter of which was used by Conway in his initial experiments).

For Test 1, five distinct ‘soups’, with each cell having 50\% probability of being alive or dead, were randomly generated, evolved for 100 generations, then had their final population taken. This was carried out on both box and toroidal grids, both with 2048 by 2048 cells overall. 

For Test 2, all of the $2^{16}$ different configurations of a binary 4 by 4 grid were tested as the initial central square of a 25 by 25 grid, and again evolved for 100 generations, both on box and torus.

For Test 3, only a 2 by 2 toroidal grid was used for each ruleset. This has 16 possible configurations, but only a few of these will give valid  $\Phi$ values. At this stage it is necessary to justify the use of such a relatively small system for testing using IIT. 

For the third test, a given ruleset first had to be conceptualised as a system of logic gates, which could then be converted into a Transition Probability Matrix (TPM). This TPM is the raw input of this ruleset into the IIT algorithm, and is an array of dimensions $n \times 2^n $, where n is the number of nodes in the system. For a square grid of side length s, n = $s^2$. This is inputted alongside the Connectivity Matrix (CM), which encodes the relevant grid structure, and has dimensions $n \times n$.

 Just as these input values grow exponentially in size as the  number of cells considered grows, the runtime of the PyPhi algorithm will also grow exponentially, going as $O(n \times 53^n)$. Furthermore, each distinct valid configuration of a set of cells - of which there are at maximum $2^n$ - will have its own distinct $\Phi$ value. 

 In practice, therefore, it was found that a maximum of n = 4, or a 2 by 2 grid, was the largest feasible sample that could be studied for each ruleset.

 \begin{figure}[h!]
\begin{center}
 \includegraphics[width=3in]{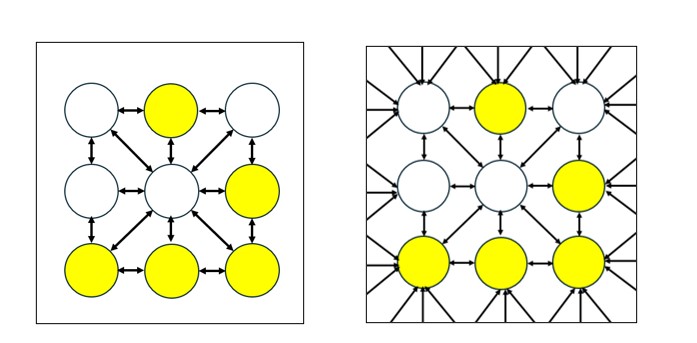} 
 \caption{Network diagrams of a box grid (left) and a toroidal grid (right). Note that cells in the right image connect to cells on the opposite side of the 'grid', producing many more interconnections. }
   \label{fig1}
\end{center}
\end{figure}
 
\section{Results}
All three tests were successfully able to select CGOL as an outlier in the LCA set, thus parametrising an intuitive sense of livingness, with very similar results on both box and torus grids for the first two tests. 

CGOL was unusual in how low entropy its final state is after evolution from a maximum entropy 50\% state, although most LCAs did tend to move away from thermodynamic equilibrium as they evolved, and it was not the lowest entropy of all states, which can be attributed partially to the fact that the minimum entropy state of a CA grid is empty, so that any predictably exploding or decaying ruleset will tend to produce a stronger anti-entropic effect by this metric. (Figure 3). This was also true on a much longer timescale. (Figure 4).

As expected, CGOL had a higher standard deviation of final population than the majority of other rulesets tested. (Figure 5). This means that given all the possible initial ‘seeds’ of different possible fillings of the 16 central cells, most rulesets evolved them in relatively predictable and convergent ways, unlike CGOL and a few other outliers, which evolved in divergent and chaotic fashion. 

 It was also found that some rulesets displayed a surprising correspondence between ash and seed density. Of the 100 rulesets, the mean population density of soup after 100 generations – averaged over five runs – was within |0.3| \% of the mean population density of a patch of 16 randomly generated cells – averaged over all its possible configurations – for 64. By way of comparison, this difference was around 7\% for CGOL. It is unclear why this is the case. 

Finally, the minimum, mean, and maximum $\Phi$ value of CGOL across its valid configurations was distinctly lower than the corresponding average value of the other LCAs tested (Figure 6). 

This is interesting, as going by a naive implementation of the algorithm as originally conceived in IIT, higher $\Phi$ would correspond to a greater 'consciousness', or in this case livingness. However, connecting back to the idea of criticality, increasing the $\Phi$ of a real brain beyond a certain point would correspond to supercriticality, and in fact would represent a state of epileptic seizure. (\cite[Hesse \& Gross. 2014]{ref-z})  This actually supports the idea that CGOL exists at a critical point within the space of LCA, neither having too high of a $\Phi$ value or too low. 

A strange result of Test 3 was that configurations that were simply rotations or mirrorings of each other would sometimes give different outputs from the IIT algorithm (Figure 6). It's unclear why this is. 

\begin{figure}[h!]
\begin{center}
 \includegraphics[width=5in]{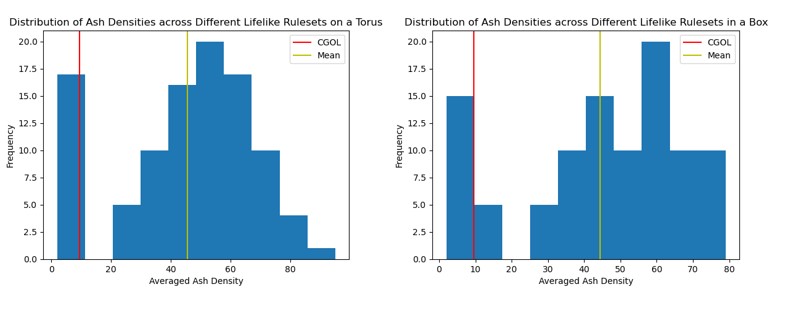} 
 \caption{For both a box and a torus grid, the final value towards which the population density of 'soups' - ash density - is left-shifted for CGOL relative to the mean of 100 other rulesets, indicating a lower entropy equilibrium.}
   \label{fig1}
\end{center}
\end{figure}

\begin{figure}[h!]
\begin{center}
 \includegraphics[width=3in]{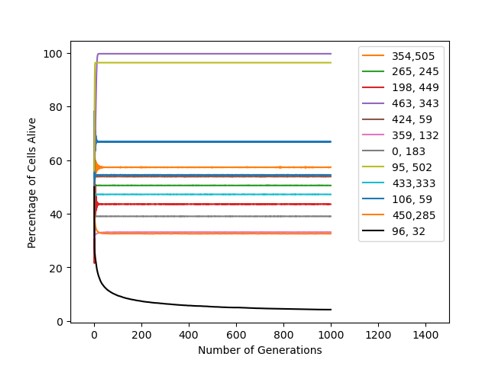} 
 \caption{A set of different rulesets are allowed to evolve from soup to ash. The lowest line is CGOL, which is clearly again moving towards a lower entropy equilibrium than its counterparts. }
   \label{fig1}
\end{center}
\end{figure}

\begin{figure}[h!]
\begin{center}
 \includegraphics[width=6in]{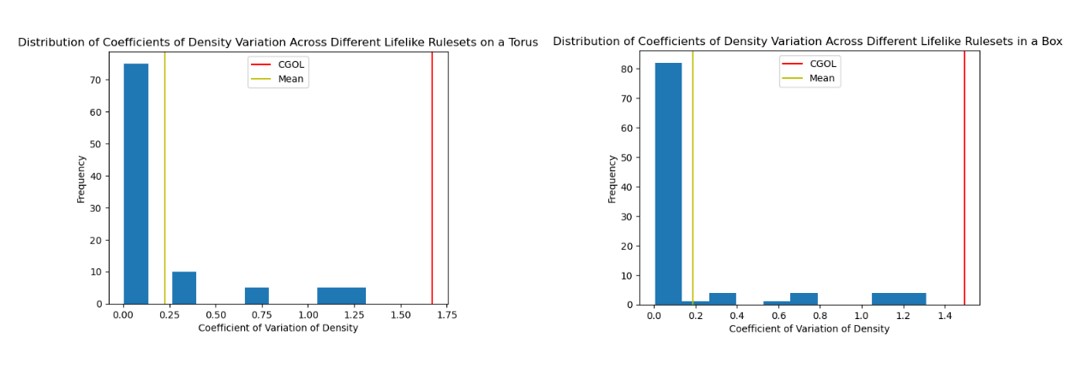} 
 \caption{The divergence of different initial configurations is much higher - right-shifted - for CGOL than for the mean of other rulesets on both a torus and box. }
   \label{fig1}
\end{center}
\end{figure}

\begin{figure}[h!]
\begin{center}
 \includegraphics[width=3in]{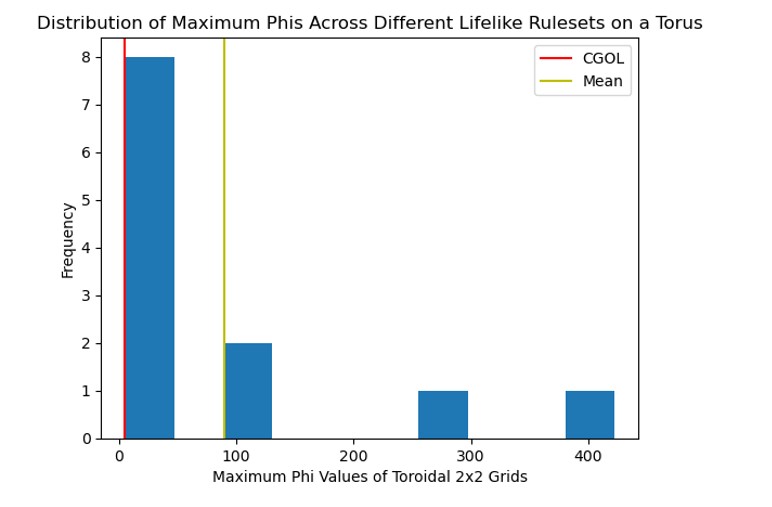} 
 \caption{The a) minimum, b) mean, and c) maximum phi values of the different valid configurations of a 2 by 2 grid were all smaller for CGOL than for most other LCAs. d) The different configurations (white = dead, black = live) of a 2 by 2 grid and their outputs from the IIT algorithm for the CGOL ruleset. }
   \label{fig1}
\end{center}
\end{figure}

\section{Discussion}
We may think of Conway’s Game of Life as simply one world, amongst many possible worlds of the same kind.

We can analogise its special place within the LCA ruleset to the special place that Earth occupies amongst known planets, or the special place that our ‘fine-tuned’ universe may theoretically occupy within a wider multiverse. (\cite[Goff. 2024]{ref25})

It is the configuration of a specific set of parameters that is most conductive to what we may call life. If the physical constants of our universe were slightly different, then life could not exist – indeed, atoms could not exist. Equally, if some currently not fully characterised set of parameters of planet Earth – stellar type, planetary distance, satellite size – were different, Earth would be a dead world. 

Lifelike cellular automata are controlled by a similar set of highly sensitive parameters, such that changing any one of the crucial 18 bits can mean the difference between a very rich and qualitatively interesting system and the kind of system that Conway dismissed as too uninteresting to bother playing with. 

Notably, unlike those analogue cases, we have very little idea how these parameters exactly relate to the behaviour of the system. We are aware of how, for example, the Earth’s position relative to the Sun affects its living systems, and so can look for planets with similar positioning. We know that altering physical constants in certain ways would result in certain kinds of universes. While Wolfram has produced a classification system for different kinds of cellular automata, even Conway was only able to find the class he was looking for by trial and error. We do not understand – and it may be unknowable – how exactly one should want to configure the 18 bits to create a non-CGOL ruleset that seems highly living, or, indeed, how to reliably produce lifelike rulesets in other cellular automaton systems. 

However, we can, in essence, identify ‘biosignatures’ that mark out habitable rulesets. Because we know that CGOL is a qualitatively lifelike system, we can hypothesise that it will have certain quantitatively lifelike traits, and then search for these traits across the space. In this study, we have identified three such abstract biosignatures – criticality, anti-entropic behaviour, and interconnectivity. The first two were definitively able to separate CGOL from similar rulesets; the third requires further study, but seems promising in this very early stage.

This is instructive both in the development of new ways to locate habitable, or inhabited, planets and worlds, and also in the use of non-biological systems as more than simply an analogue to a living system. CGOL is an excellent example of many different physical processes, and indeed, it is an excellent example of a kind of non-biological life in its own right. Its similarities are more than just surface deep; it has quantitative properties in common, not just a qualitative resemblance. 

Further research is needed to characterise more of the lifelike ruleset of cellular automata, as well as to run similar tests on other cellular automata outside of this ruleset. Other rich and interesting rulesets may be found in this way, and the underlying markers of such systems understood more clearly. Furthermore, the tests for criticality and integration in addition to far-from-equilibrium behaviours can be applied to a wider variety of systems than relatively simple cellular automata, which stand here as a case in point to testify to their usefulness.

\section{Acknowledgements}
The authors would like to thank Gaelan Steele for her commentary and support during the early stages of this work, which had a great impact on inspiring and improving its path to its final form. 
The authors would like to thank the Life community for its tireless categorisation and study of the game and its variants for more than fifty years since its inception, especially Achim Flammenkamp, whose extensive experiments with it formed a core part of this publication. 
This paper would also of course be completely impossible without the work of John H Conway, and would be greatly lacking were it not for his detailed and insightful account of the process of creation of his work in cellular automata. May he rest in peace. 

\section{Appendix I}
The Flammenkamp value of the attractor of soup population density for the CGOL ruleset - 2.87115\% - is strangely close to Euler's number, e. This is not the only similar value for CGOL that is close to this constant. 

If we start with a 50\% population density grid and run one iteration of Conway’s Game of Life, the population density will generally be around 27\%. This experimental result can also be predicted using this equation:

$P =\frac{1}{2} B(3,8,\frac{1}{2})+\frac{1}{2} B(2,8,\frac{1}{2})+\frac{1}{2} B(3,8,\frac{1}{2})$

$P =\frac{1}{2} (0.21875)+\frac{1}{2}(0.10938)+\frac{1}{2} (0.21875)=0.27344$

Recursively applying this equation with a starting value of 0.5 for CGOL gives a limiting value of approximately $\frac{1}{e}$.

\section{Appendix II}

\begin{table}[h!]
    \centering
    \begin{tabular}{cccccccccc}
         448/131
&  221/151&  333/329&  398/165&  153/487&  080/280&  323/000&  244/130&  292/361& 509/212
\\
         286/117
&  234/381&  002/484&  369/366&  347/116&  480/505&  053/364&  132/413&  404/024& 477/354
\\
         411/168
&  203/496&  405/247&  218/059&  098/413&  242/341&  074/170&  500/368&  096/139& 325/193
\\
         438/370
&  148/278&  066/044&  253/355&  174/493&  494/194&  134/122&  119/237&  012/244& 289/185
\\
         453/222
&  358/008&  069/440&  051/481&  036/369&  438/040&  202/411&  234/500&  483/389& 187/478
\\
         338/404
&  116/491&  450/322&  266/128&  446/449&  150/337&  376/054&  160/192&  183/374& 331/314
\\
         161/043
&  454/031&  387/367&  317/008&  302/403&  209/486&  312/113&  052/268&  261/323& 370/443
\\
         191/163
&  267/315&  325/077&  162/274&  473/160&  146/127&  361/250&  347/476&  293/350& 480/474
\\
         426/167
&  186/307&  071/104&  292/148&  181/411&  020/296&  308/456&  501/212&  287/279& 446/388
\\
         134/028
&  342/320&  095/121&  495/188&  478/281&  229/273&  433/387&  479/485&  484/024& 153/350
\\
    \end{tabular}
    \caption{Rulesets used in this paper.}
    \label{tab:my_label}
\end{table}


\begin{thebibliography}{}
\bibitem[Alstrøm \& Leão. 1994]{ref10}
 Alstrøm P. \& Leão J., “Self-organized criticality in the ‘‘game of Life’’,” Physical Review E, vol. 49, no. 4, 1994. 
 \bibitem[Azua-Bustos \& Martínez. 2013]{ref-a}
 Azua-Bustos A., Vega-Martínez C. The potential for detecting ‘life as we don’t know it’ by fractal complexity analysis. 2013 International Journal of Astrobiology. 12(4), 314.

\bibitem[Bak et al. 1989]{ref11}
 Bak, P., Chen K., \& Creutz M., “Self-organized criticality in the 'Game of Life",” Nature, vol. 342, pp. 780-782, 1989.
 
\bibitem[Berkelamp et al. 2004]{ref6} 
 Berkelamp, E., Conway, J.H., \&  Guy R., Winning Ways for your Mathematical Plays, 2nd ed., AK Peters Ltd, 2004.
 
\bibitem[Cleland \& Cheba. 2010]{ref1}
Cleland, C. \& Chyba, C. (2010). Does 'life' have a definition?. Planets and Life: The Emerging Science of Astrobiology. 119-131. 10.1017/CBO9780511730191.032.

\bibitem[Cornish-Bowden \& Cárdenas. 2019]{ref2}
Cornish-Bowden, A. \& Cárdenas, M. (2019). Contrasting Theories of Life: Historical Context, Current Theories. In search of an ideal theory. Biosystems. 188. 104063. 10.1016/j.biosystems.2019.104063. 

\bibitem[Doerig et al. 2019]{ref22}
 Doerig,A., Schurger,A., Hess, K., \& Herzog M.H, “The unfolding argument: Why IIT and other causal structure theories cannot explain consciousness,” Consciousness and Cognition, vol. 72, pp. 49-59, 2019. 

\bibitem[Eppstein. 2010]{ref19}
 Eppstein,D., “Growth and decay in life-like cellular automata,” in Game of Life Cellular Automata, A. Adamatzky, Ed., Springer, 2010, pp. 71-98.

\bibitem[Flammenkamp. 2004]{ref20}
 Flammenkamp,A., “Most seen naturally occurring ash objects in Game of Life,” 2004. [Online]. Available: https://wwwhomes.uni-bielefeld.de/achim/freq\_top\_life.html. [Accessed 25th May 2024].

\bibitem[Gardner. 1970]{ref4}
 Gardner, M. “The fantastic combinations of John Conway's new solitaire game "life",” Scientific American, Mathematical Games, vol. 223, no. 4, pp. 120-123, 1970.

 \bibitem[Goff. 2024]{ref25}
Goff, P. "Is the fine-tuning evidence for a multiverse?". Synthese 204, 3 (2024). https://doi.org/10.1007/s11229-024-04621-z

\bibitem[Grosu et al. 2023]{ref-b}
 Grosu, G.F., Hopp, A.V., Moca, V.V., Bârzan, H., Ciuparu, A., Ercsey-Ravasz, M., Winkel, M., Linde, H., Mureșan, R.C The fractal brain: scale-invariance in structure and dynamics. 2023 Cerebral Cortex, 33(8), 4574.

 \bibitem[Hesse \& Gross. 2014]{ref-z}
 Hesse J, Gross T. Self-organized criticality as a fundamental property of neural systems. Front Syst Neurosci. 2014 Sep 23;8:166. 

 \bibitem[Izhikevich et al. 2015]{ref5}
 Izhikevich,E. M.  Conway, J. H. \%  Seth, A. “Game of Life,” 2015. [Online]. Available: http://www.scholarpedia.org/article/Game\_of\_Life. [Accessed 26th May 2024].

\bibitem[Landauer. 1961]{ref17}
 Landauer,R. “Irreversibility and Heat Generation in the Computing Process,” IBM Journal of Research and Development, vol. 5, no. 3, pp. 183-191, 1961. 

\bibitem[Macii \& Poncino. 1996]{ref16}
 Macii, E \&  Poncino,M. “Exact Computation of the Entropy of a Logic Circuit,” Proceedings of the Sixth Great Lakes Symposium on VLSI, 1996. 

\bibitem[Mayner et al. 2018]{ref24} 
 Mayner, W., Marshall, W., Albantakis, L., Findlay, G., Marchman, R., \& Tononi,G., “PyPhi: A toolbox for integrated information theory,” PLOS Computational Biology, vol. 14, no. 7, 2018.

\bibitem[McKay. 2004]{ref-c}
 McKay, C.P. What is life--and how do we search for it in other worlds? 2004 PLoS Biol. 2(9), E302.

\bibitem[Merker et al. 2022]{ref23} 
 Merker, B., Williford, K., \& Rudrauf,D., “The integrated information theory of consciousness: a case of mistaken identity,” Behavioural and brain sciences , 2022. 

\bibitem[Schiff. 2011]{ref7}
 Schiff J., Cellular Automata: A Discrete View of the World, Wiley \& Sons, Inc, 2011. 

\bibitem[Schrodinger. 1944]{ref15}
 Schrödinger,E. What Is Life? The Physical Aspect of the Living Cell, Cambridge University Press, 1944.

\bibitem[Shannon. 1948]{ref18}
 Shannon,C.E., “A Mathematical Theory of Communication,” Bell Systems Technical Journal, vol. 27, pp. 379-423, 1948.

\bibitem[Tian et al. 2022]{ref8}
 Tian, Y., Tan,Z., Hou,H., Li, G., Cheng,A., Qiu,Y., Weng, K., Chen, C., \& Sun, P., “Theoretical foundations of studying criticality in the brain,” Network Neuroscience, vol. 6, no. 4, pp. 1148-1185, 2022. 

\bibitem[Tononi et al. 2016]{ref21}
 Tononi,G., M. Boly, M. Massimini and C. Koch, “Integrated Information Theory: From consciousness to its physical substrate,” Nature Reviews Neuroscience, 2016. 

\bibitem[Turing. 1937]{ref14}
 Turing,A.M, “On Computable Numbers, with an Application to the Entscheidungsproblem,” Proceedings of the London Mathematical Society, Vols. s2-42, no. 1, pp. 230-265, 1937.

\bibitem[von Neumann \& Burks. 1966]{ref3}
 von Neumann, J. \& Burks,A. W., Theory of Self-Reproducing Automata., University of Illinois Press, 1966.


\bibitem[Wolfram. 1984]{ref12}
 Wolfram,S., “Universality and complexity in cellular automata,” Physica D: Nonlinear Phenomena, vol. 10, no. 1-2, pp. 1-35, 1984. 

\bibitem[Wolfram. 2002]{ref13}
 Wolfram,S., A New Kind of Science, Wolfram Media, 2002. 

\end{thebibliography}
\end{document}